\documentclass[
aps,
pre, 
amsmath,amssymb,
reprint,%
]{revtex4-1}
\usepackage{graphicx}
\usepackage{dcolumn}
\usepackage{bm}

\usepackage{amsmath}	
\begin{document}

\newcommand{\diff}[2]{\frac{d#1}{d#2}}
\newcommand{\pdiff}[2]{\frac{\partial #1}{\partial #2}}
\newcommand{\fdiff}[2]{\frac{\delta #1}{\delta #2}}
\newcommand{\bx}{\bm{x}}
\newcommand{\bq}{\bm{q}}
\newcommand{\br}{\bm{r}}
\newcommand{\bu}{\bm{u}}
\newcommand{\by}{\bm{y}}
\newcommand{\bY}{\bm{Y}}
\newcommand{\bF}{\bm{F}}
\newcommand{\new}{\nonumber\\}
\newcommand{\abs}[1]{\left|#1\right|}
\newcommand{\tr}{{\rm Tr}}
\newcommand{\HH}{{\mathcal H}}
\newcommand{\II}{{\mathcal I}}
\newcommand{\WW}{{\mathcal W}}
\newcommand{\OO}{{\mathcal O}}
\newcommand{\ave}[1]{\left\langle #1 \right\rangle}
\newcommand{\im}{{\rm Im}}
\newcommand{\re}{{\rm Re}}
\newcommand{\ke}{k_{\rm eff}}
\newcommand{\ipr}{{\rm IPR}}
\newcommand{\TT}{\mathcal{T}}
\newcommand{\tg}{\tilde{\Gamma}}

\preprint{AIP/123-QED}

\title{Correlated Noise and Critical Dimensions}

\author{Harukuni Ikeda}
 \email{harukuni.ikeda@gakushuin.ac.jp}
\affiliation{ Department of Physics, Gakushuin University, 1-5-1 Mejiro,
Toshima-ku, Tokyo 171-8588, Japan}

\date{\today}

\begin{abstract}
In equilibrium, the Mermin-Wagner theorem prohibits the continuous
symmetry breaking for all dimensions $d\leq 2$.  In this work, we
discuss that this limitation can be circumvented in non-equilibrium
systems driven by the spatio-temporally long-range anticorrelated
noise. We first compute the lower and upper critical dimensions of the
$O(n)$ model driven by the spatio-temporally correlated noise by means of
the dimensional analysis. Next, we consider the spherical model, which
corresponds to the large $n$ limit of the $O(n)$ model and allows us to
compute the critical dimensions and critical exponents,
analytically. Both results suggest that the critical dimensions increase
when the noise is positively correlated in space and time, and decrease
when anticorrelated. We also report that the spherical model with the
correlated noise shows the hyperuniformity and giant number fluctuation
even well above the critical point.
\end{abstract}

\maketitle

\section{Introduction}

The minimum dimension required for a phase transition to occur is known
as the lower critical dimension $d_l$~\cite{nishimori2010elements}. For
systems with quenched randomness, Imry and Ma predicted that the lower
critical dimension is $d_l=2$ for a discrete order parameter and $d_l=4$
for a continuous order parameter~\cite{imry1975}. Recent studies have
reported that $d_l$ can be reduced by introducing anticorrelation to the
quenched randomness. For example, in Ref.~\cite{schwartz1993}, the
authors studied the random field Ising model with the anticorrelated
random field and showed that the ordered phase arises on the ground
state even in $d=2$. Ref.~\cite{janke1995two} reported a first-order
transition of the Potts model on a random Voronoi lattice in $d=2$.  The
authors in Ref.~\cite{barghathi2014} argued that this is a consequence
of the strong anticorrelation in the coordination number of the random
Voronoi lattices, which reduces the lower critical dimension. Similar
anticorrelation also appears in the Ising model in the aperiodic field,
which stabilizes the ferromagnetic ground state even in
$d=1$~\cite{luck1987frustration,sire1993ising}.

For equilibrium systems without quenched randomness, the Mermin-Wagner
theorem claims that the lower-critical dimension of the continuous
symmetry breaking is $d_l=2$~\cite{mermin1969}. However, for
out-of-equilibrium systems, the continuous symmetry breaking can occur
even $d\leq 2$; some examples include the XY model driven by anisotropic
noise~\cite{xy1995,reichl2010}, $O(n)$ model driven by
shear~\cite{corberi2002,nakano2021}, Vicsek
model~\cite{vicsek1995,toner1995}, and so
on~\cite{loos2022long,dadhichi2020}.  A recent numerical study suggests
that so-called hyperuniform states of
matter~\cite{torquato2018hyperuniform} may potentially be added to the
list above~\cite{leonardo2023}. The hyperuniform states of matter are
characterized by the anomalous suppression of the density fluctuation on
a large scale, which leads to the vanishing of the static structure
factor $S(q)$ in the limit of the small wave number
$\lim_{q\to}S(q)=0$~\cite{torquato2018hyperuniform}. This property is
referred to as the hyperuniformity (HU), which was first introduced for
density fluctuation, but later has been extended to fluctuations of more
general quantities, such as spin variables and vector
fields~\cite{torquato2018hyperuniform}. In Ref.~\cite{leonardo2023},
Galliano {\it et al.} proposed and numerically demonstrated that the
suppression of the density fluctuation also reduces $d_l$, leading to
the perfect crystalline phase even in $d=2$.

The HU is widely observed in various systems such as Vycor
glass~\cite{crossley1991image}, periodically driven
emulsions~\cite{weijs2015emergent}, chiral active
matter~\cite{huang2021circular,hyperchiral2022}, and so
on~\cite{torquato2018hyperuniform}.  While no unified theory has been
found that can comprehensively explain the HU observed in all these
systems, if it exists, Hexner and Levine proposed that the HU can
universally appear for systems having certain
symmetries~\cite{hexner2017noise}. In Ref.~\cite{hexner2017noise}, they
derived a Langevin equation for a system conserving the total number of
particles and center of mass.  The effective noise of the resultant
Langevin equation has spatial anticorrelation, which leads to the
HU~\cite{hexner2017noise,lei2019hydrodynamics}. Other examples showing
the HU are chiral active
matter~\cite{huang2021circular,hyperchiral2022}, where the periodic
nature of the driving force leads to the temporal anticorrelation of the
effective noise, which results in the
HU~\cite{kuroda2023microscopic,ikeda2023does}. These results suggest
that spatial or temporal anticorrelation of the noise leads to HU, which
may reduce $d_l$.

Based on the above observations, it is tempting to conjecture that the
anticorrelation of the noise or quenched randomness generally reduces
the lower critical dimension. Our first goal is to test this conjecture.
The second goal is to investigate the effects of the long-range temporal
correlation of the noise on critical phenomena. The effects of the
temporally correlated noise on a single particle have been investigated
significantly in the context of anomalous
diffusion~\cite{bouchaud1990anomalous,eliazar2009unified}. However, its
effects on many-body systems, in particular near the critical point,
have not been investigated sufficiently before. For those goals, here we
investigate the effects of the spatiotemporally correlated noise on the
second-order phase transition by using the $O(n)$ and spherical
models. As discussed in the following paragraph, the noise encompasses
the time-independent random field and equilibrium white noise in certain
limits.

For concreteness, we consider model-A and B
dynamics~\cite{chaikin1995principles,nishimori2010elements} with the
correlated noise $\xi(\bx,t)$ of zero mean and variance
\begin{align}
\ave{\xi(\bx,t)\xi(\bx',t')} =2TD(\bx-\bx',t-t'),
\end{align}
where $D(\bx,t)$ represents the spatiotemporal correlation of the
noise. The Fourier transform of $D(\bx,t)$ w.r.t. $\bx$ and $t$ is given
by
\begin{align}
D(\bq,\omega)=\abs{\bq}^{-2\rho}\abs{\omega}^{-2\theta},\label{104934_23Feb23}
\end{align}
where $\bq$ denotes the wave vector, and $\omega$ denotes the frequency.
The same correlation function has been considered in previous works to
investigate the effects of the long-range spatio-temporal correlation on
the Kardar-Parisi-Zhang (KPZ)
equation~\cite{burger1989,janssen1999exact,kpz2004,kpz2019}. When
$\rho=\theta=0$, the noise can be identified with the white noise in
equilibrium.  The positive values of $\rho$ and $\theta$ represent the
positive power-law correlation in the real space: $D(\bx,t)\sim
\abs{\bx}^{2\rho-d}\abs{t}^{2\theta-1}$, where $d$ denotes the spatial
dimension. In the limit $\theta\to 1/2$, the noise correlation does not
decay with time, and thus the noise can be identified with the quenched
random field. The negative values of $\rho$ and $\theta$ imply the
existence of the anticorrelation because $D(\bq=0,\omega=0)=\int d\bx
\int dt D(\bx,t)=0$. Therefore, the model can smoothly connect the white
noise ($\rho=\theta=0$), quenched randomness ($\theta\to1/2$),
positively correlated noise ($\theta>0$, $\rho>0$), and anticorrelated
noise ($\theta<0$, $\rho<0$) by changing $\rho$ and $\theta$. In this
work, we show that the positive correlation increases the lower and
upper critical dimensions $d_l$ and $d_u$, and the anticorrelation
reduces $d_l$ and $d_u$.  


The structure of the paper is as follows. In Sec.~\ref{173631_21Feb23},
we investigate the $O(n)$ model driven by the model-A dynamics with the
correlated noise by means of the dimensional analysis.
Sec.~\ref{145155_19Mar23}, we investigate the spherical model, which
corresponds to the $n\to\infty$ limit of the $O(n)$ model and allows us
to calculate the critical dimensions
analytically~\cite{nishimori2010elements}. We also discuss that the
positive correlation of the noise induces the giant number fluctuation
(GNF), {\it i.e.}, the anomalous enhancement of the fluctuation even far
above the critical point. On the contrary, the anticorrelation of the
noise suppresses the fluctuation and induces the HU. In
Sec.~\ref{145218_19Mar23}, we discuss the behavior of the conserved
order parameter driven by the model-B dynamics with the correlated
noise. In Sec.~\ref{130256_22Feb23}, we summarize the work.

\section{Dimensional analysis}
\label{173631_21Feb23} Here we derive the upper and lower critical
dimensions of the $O(n)$ model driven by the model-A dynamics with
the correlated noise. 

\subsection{Model}
Let $\vec{\phi}=\{\phi_1,\cdots,\phi_n\}$ be a non-conserved
$n$-component order parameter. We assume that the time evolution of
$\phi_a(\bx,t)$ follows the model-A dynamics~\cite{chaikin1995principles}:
\begin{align}
\pdiff{\phi_a(\bx,t)}{t} = -\Gamma\fdiff{F[\vec{\phi}]}{\phi_a(\bx,t)} + \xi_a(\bx,t),
 \label{141313_15Feb23}
\end{align}
where $\Gamma$ denotes the damping coefficient, and $\xi_a$ denotes the
noise. $F[\vec{\phi}]$ denotes the free energy of the $O(n)$
model~\cite{nishimori2010elements}:
\begin{align}
F[\vec{\phi}] = \int d\bx \left[
 \frac{\sum_{a=1}^n \nabla\phi_a\cdot\nabla\phi_a}{2}+
 \frac{\varepsilon|\vec{\phi}|^2}{2}
 +\frac{g|\vec{\phi}|^4}{4} \right],
 \label{153105_27Feb23}
\end{align}
where
$|\vec{\phi}|^2=\sum_{a=1}^n\phi_a^2$, $\varepsilon$ denotes
the linear distance to the transition point, and $g$ denotes the
strength of the non-linear term. The mean and variance of the noise
$\xi_a(\bx,t)$ are
\begin{align}
&\ave{\xi_a(\bx,t)} =0,\new 
&\ave{\xi_a(\bx,t)\xi_b(\bx',t')} =2T\delta_{ab}\Gamma D(\bx-\bx',t-t'),
\end{align}
where $D(\bx,t)$ represents the correlation of the noise. We assume that
the correlation in the Fourier space is written
as~\cite{burger1989,janssen1999exact,kpz2004,kpz2019}
\begin{align}
D(\bq,\omega) = \abs{\bq}^{-2\rho}\abs{\omega}^{-2\theta}.\label{094611_17Mar23}
\end{align}
To ensure the existence of the Fourier transform of $D(\bq,\omega)$, the
values of $\rho$ and $\theta$ are constrained to $\rho<d/2$ and
$\theta<1/2$, and one should introduce the high-frequency cutt-off for
$\theta\leq -1/2$. The noise can be generated, for instance, by
integrating uncorrelated white noise $\eta_a(\bx,t)$ with a proper kernel
$K(\bx,t)$:
\begin{align}
\xi_a(\bx,t) = \int_{-\infty}^{\infty}dt\int d\bx K(\bx-\bx',t-t')\eta_a(\bx',t'),
\end{align}
where $K(\bx,t)$ satisfies $K(\bx,t)=0$ for $t<0$ and
$\abs{K(\bq,\omega)}\sim\abs{\bq}^{-\rho}\abs{\omega}^{-\theta}$ in the
Fourier space~\cite{burger1989}. Another way to generate the correlated
noise would be to perturb periodic patterns, which naturally leads to
the anticorrelation ($\rho<0$ and $\theta<0$)~\cite{kim2018}, see also
Sec.~\ref{imper} for a related discussion. The model satisfies the
fluctuation-dissipation theorem only when
$\rho=\theta=0$~\cite{zwanzig2001nonequilibrium}. For $\rho\neq 0$ or
$\theta\neq 0$, on the contrary, the model violates the detailed
balance, and thus the steady-state distribution is not given by the
Maxwell-Boltzmann distribution~\footnote{ To satisfy the detailed
balance, the left hand side of Eq.~(\ref{141313_15Feb23}) should be
replaced as
\begin{align}
\pdiff{\phi_a(\bx,t)}{t}\to \int^{t}_{-\infty}dt'\int d\bx'D(t-t',\bx-\bx') \pdiff{\phi_a(\bx',t')}{t'},
\end{align}
{\it i.e.}, one should consider the (overdamped) generalized Langevin
equation ~\cite{zwanzig2001nonequilibrium,bonart2012critical}.}.

\subsection{Critical dimensions}

From Eqs.~(\ref{141313_15Feb23}) and (\ref{153105_27Feb23}), we get
\begin{align}
 \dot{\phi}_a= -\Gamma (-\nabla^2\phi_a+\varepsilon\phi_a + g\phi_a|\vec{\phi}|^2)+\xi_a
 \label{151905_20Mar23}
\end{align}
Now we consider the following scaling transformations: $x\to bx$, $t\to
b^{z_t}t$, $\phi_a\to b^{z_\phi}\phi_a$, $g\to
b^{z_g}g$~\cite{nishimori2010elements}. To calculate the scaling
dimension of the noise, we observe the fluctuation induced by the noise
in a ($d+1$)-dimensional Euclidean space $[0,l]^d\times
[0,t]$~\cite{torquato2018hyperuniform}:
\begin{align}
\sigma(l,t)^2 \equiv \ave{\left(\int_{\bx'\in [0,l]^d}
 \hspace{-5mm}d\bx' \int_0^t dt'\xi_a(\bx',t')\right)^2}.
\end{align}
The asymptotic behavior for $l\gg 1$ and $t\gg 1$ is
\begin{align}
 \sigma(l,t)^2 \sim
 (c_1t^{1+2\theta}+c_2t^0)\left(c_3l^{d+2\rho} + c_4l^{d-1}\right),\label{151754_20Mar23}
\end{align}
where $c_i$ denotes a constant, and $c_2t^0, c_4l^{d-1}$ account for the
surface 
contributions~\cite{nattermann1998theory,torquato2018hyperuniform}.
Eq.~(\ref{151754_20Mar23}) implies $\xi(\bx,t)\to
b^{z_t(2\theta'-1)/2}b^{(2\rho'-d)/2}\xi(\bx,t)$, where
\begin{align}
&\rho'=\max[\rho,-1/2],\ \theta'=\max[\theta,-1/2].\label{170358_1Aug23}
\end{align}
Assuming the scaling invariance of the dynamics
Eq.~(\ref{151905_20Mar23}), we
get~\cite{nishimori2010elements,maggi2022critical}
\begin{align}
&z_t = 2,\new
&z_g = -2z_\phi-z_t,\new 
&z_\phi =1 + \frac{2\rho'-d}{2}+2\theta'.\label{144150_21Sep23}
\end{align}
The simplest way to calculate the lower critical dimension $d_l$ is to
observe the fluctuation of the order parameter:
\begin{align}
\ave{\delta\phi_a^2}\sim b^{2z_{\phi}}.
\end{align} 
To ensure the stability of the ordered phase, $z_\phi$ must be negative;
otherwise, the fluctuation of the order parameter diverges in the
thermodynamic limit $b\to\infty$, which destroys the long-range
order. Therefore, the lower-critical dimension can be determined by
setting $z_\phi=0$, leading to
\begin{align}
d_l =
2+2\rho'+4\theta'.
\label{024537_17Mar23} 
\end{align}
Eq.~(\ref{144150_21Sep23}) implies that the coupling of the non-linear
term $g$ scales as $g'=b^{-z_g}g$ under the scaling transformation
$x'=b^{-1}x$~\cite{nishimori2010elements,maggi2022critical}. When
$z_g>0$, the non-linear term is irrelevant, and vice versa. Therefore,
the upper critical dimension is obtained by setting $z_g=0$, leading to
\begin{align}
d_u =
4+2\rho'+4\theta'.
\end{align}
When $\rho=\theta=0$, we get $d_l=2$ and $d_u=4$, which are consistent
with the standard $O(n)$ model in
equilibrium~\cite{nishimori2010elements}.  When $\rho\neq 0$ or
$\theta\neq 0$ on the contrary, the system reaches the non-equilibrium
steady state because the noise does not satisfy the detailed balance.
In this case, the positive correlation of the noise ($\rho>0$,
$\theta>0$) increases the critical dimensions, $d_l$ and $d_u$, and the
anticorrelation reduces $d_l$ and $d_u$.

\subsection{Correlated random Field}
In the limit $\theta\to 1/2$, the noise correlation does not decay with
time $D(\bx,t)\sim t^{2\theta-1}\to t^0$, and thus the noise can be
identified with the correlated random field. In this case, we get
\begin{align}
d_l^{\rm RF} =
4+2\rho'.
\label{024614_17Mar23} 
\end{align}
It would be instructive to compare the above result with the standard
Imry-Ma argument for the lower critical
dimension~\cite{imry1975,schwartz1993,nattermann1998theory}. In a domain
of linear size $l$, the typical fluctuation induced by the correlated
random field is $\sigma^2\equiv \ave{\left(\int_{\bx\in [0,l]^d} d\bx
h\right)^2}\sim c_1l^{d+2\rho}+
c_2l^{d-1}$~\cite{schwartz1993,nattermann1998theory}. The domain wall
energy is $\gamma \sim l^{d-1}$ for a discrete order parameter, and
$\gamma \sim l^{d-2}$ for a continuous order
parameter~\cite{imry1975}. When $\sigma\gg \gamma$, the fluctuation of
the random field destroys the ordered phase, and vice versa. Therefore,
on the lower critical dimension $d_l$, $\sigma\sim \gamma$, leading
to~\cite{schwartz1993}
\begin{align}
 d_l^{\rm I.M.} =
\begin{cases}
 2+2\rho' & ({\rm discrete}),\\ 
 4+2\rho' & ({\rm continuous}).
\end{cases} 
 \label{031100_17Mar23}
\end{align}
The result for a continuous order parameter is consistent with that of the
dimensional analysis Eq.~(\ref{024614_17Mar23}).

\section{Spherical model}
\label{145155_19Mar23} Due to the non-linear term in the free-energy
Eq.~(\ref{153105_27Feb23}), the $O(n)$ model can not be solved
analytically. Here we instead consider a solvable model: the spherical
model, which corresponds to the $n\to\infty$ limit of the $O(n)$
model~\cite{henkel2010non,nishimori2010elements}.



\subsection{Model}
The effective free energy of the model is
\begin{align}
F[\phi] = \int d\bx \left[
 \frac{(\nabla\phi)^2}{2}+
 \frac{\mu\phi^2}{2}\right],\label{032725_17Mar23}
\end{align}
where $\mu$ denotes the Lagrange multiplier to impose the spherical
constraint~\cite{berlin1952spherical}:
\begin{align}
\int d\bx \ave{\phi(\bx)^2} = N.\label{110847_31Jul23}
\end{align}
We impose the spherical constraint for the mean value.  One can, in
principle, consider a rigid constraint $\int d\bx \phi(\bx)^2=N$,
instead of Eq.~(\ref{110847_31Jul23}). The dynamics with the rigid
constraint has some non-linear terms that make the model difficult to
solve analytically. This paper only consider the constraint for the
mean-value Eq.~(\ref{110847_31Jul23}).

For $\rho=\theta=0$, the
steady-state distribution is given by the Boltzmann distribution, and
thus one does not need to solve the dynamical equation. In this case, 
the two-point correlation of the spherical model
with the constraint Eq.~(\ref{110847_31Jul23}) agrees with that of the
rigid constraint above the critical temperature $T_c$, but is inconsistent
below $T_c$~\cite{kac1977correlation}. So hereafter, we only focus on the
behavior above $T_c$.

\subsection{Steady-state solution}
\label{123819_22Feb23}

By substituting Eq.~(\ref{032725_17Mar23}) into 
Eq.~(\ref{141313_15Feb23}) we get a linear
differential equation:
\begin{align}
\dot{\phi} = -\Gamma (-\nabla^2\phi+\mu\phi) + \xi.
\end{align}
This can be easily solved in the Fourier space:
\begin{align}
\phi(\bq,\omega) = \frac{\xi(\bq,\omega)}{i\omega + \Gamma (q^2+\mu)},
\end{align}
where
\begin{align}
\mathcal{O}(\bq,\omega) = \int dt\int d\bx e^{-i\bq\cdot\bx-i\omega t}\mathcal{O}(\bx,t).
\end{align}
The two-point correlation is calculated as 
\begin{align}
\ave{\phi(\bq,\omega)\phi(\bq',\omega')}
&= (2\pi)^{d+1}\delta(\bq+\bq')\delta(\omega+\omega')S(q,\omega),
\end{align} 
where 
\begin{align}
S(q,\omega) 
&\equiv \int dt \int d\bx e^{i\bq\cdot\bx+i\omega t}\ave{\phi(\bx,t)\phi(0,0)}\new 
&=\frac{2T\Gamma D(\bq,\omega)}{\omega^2 + \Gamma^2 (k q^2+\mu)^2}.\label{163245_15Feb23}
\end{align}

\subsection{Correlation length and relaxation time}

Since we are interested in the critical behaviors in large
spatiotemporal scales, here we analyze the scaling behavior of the
correlation function for $\abs{\bq}\ll 1$ and $\omega\ll 1$. After some
manipulations, we get
\begin{align}
S(q,\omega) =T\mu^{-2-\rho-2\theta}\mathcal{S}(\mu^{-1/2}q,\mu^{-1}\omega),\label{101344_16Feb23}
\end{align}
where
\begin{align}
\mathcal{S}(x,y) = \frac{2\Gamma x^{-2\rho}y^{-2\theta}}{y^2+\Gamma^2(x^2+1)^2}.
\end{align}
The scaling Eq.~(\ref{101344_16Feb23}) implies that the correlation length $\xi$
and relaxation time $\tau$ behave as 
\begin{align}
&\xi \sim \mu^{-1/2},\
\tau \sim \xi^z,\label{160435_18Feb23}
\end{align}
with the dynamic critical exponent
\begin{align}
z=2.\label{165020_21Feb23}
\end{align}
The correlation length and relaxation time diverge in the limit $\mu\to
0$, meaning that $\mu=0$ defines the critical point.

\subsection{Static structure factor}
\begin{figure}[t]
\begin{center}
\includegraphics[width=9cm]{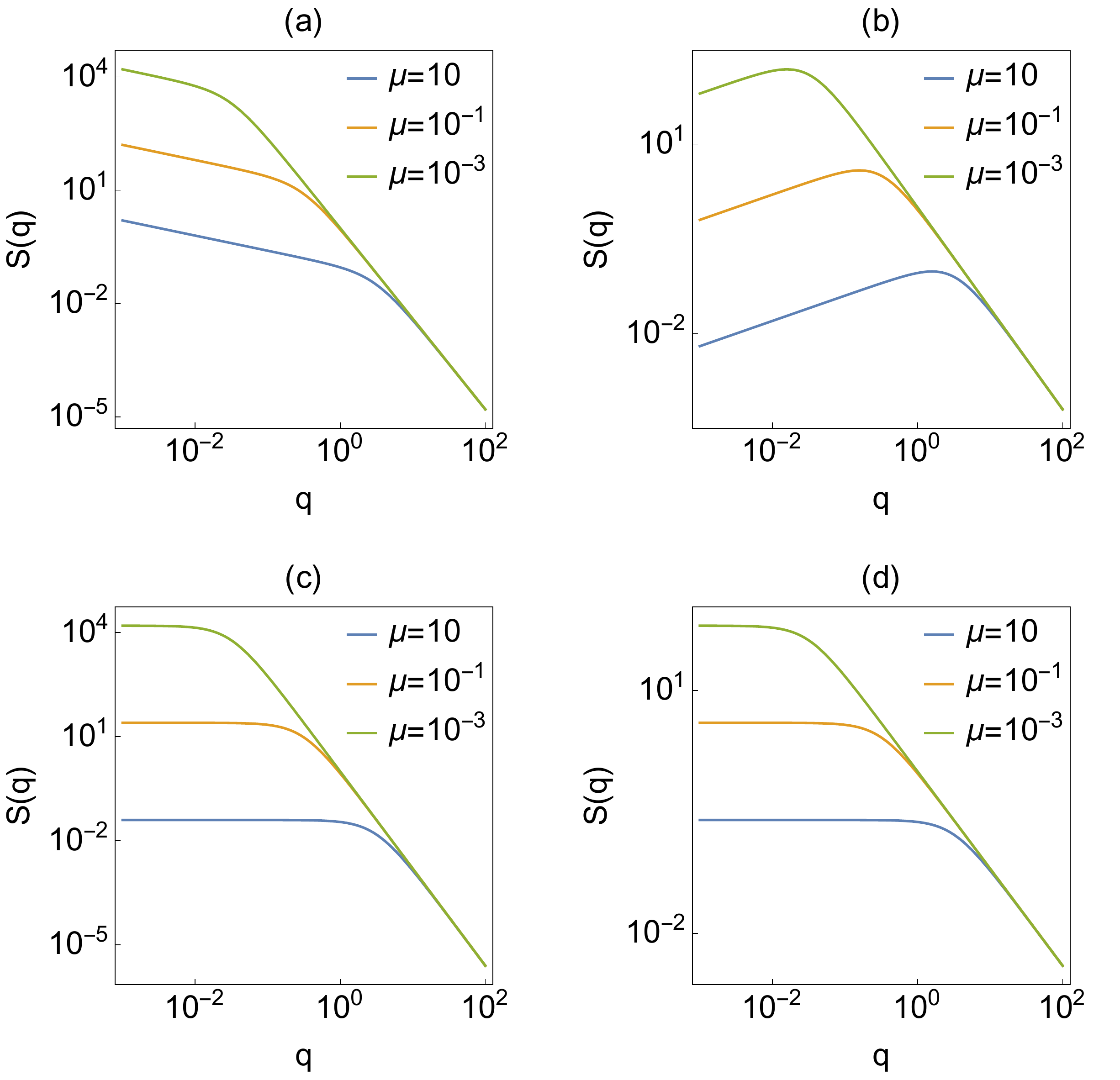} \caption{Typical behaviors of $S(q)$
of spherical model for model-A: (a) $S(q)$ for positive spatial
correlation ($\rho=0.2, \theta=0$). $S(q)$ diverges in the limit of
small $q$ even far from the transition point, {\it i.e.}, the GNF
appears. (b) $S(q)$ for negative spatial correlation ($\rho=-0.2,
\theta=0$). $S(q)$ vanishes in the limit of small $q$, {\it i.e.}, the
HU appears.  (c) $S(q)$ for positive temporal correlation ($\rho=0,
\theta=0.2$). (d) $S(q)$ for negative temporal correlation ($\rho=0,
\theta=-0.2$). The temporal correlation alone can not induce the GNF or
HU. For simplicity, here we set $A=T=1$.}\label{091838_2Aug23}
\end{center}
\end{figure}

The static structure factor $S(\bq)$ is calculated as
\begin{align}
S(\bq) = \frac{1}{2\pi}\int_{-\infty}^{\infty} d\omega S(\bq,\omega)
= \frac{ATq^{-2\rho}}{(q^2+\mu)^{1+2\theta}}\label{sq}
\end{align}
where
\begin{align}
A = \frac{1}{\pi}\int_{-\infty}^\infty \frac{\abs{x}^{-2\theta}dx}{x^2+1}
= \sec(\pi\theta).
\end{align}
Note that this integral diverges when $\theta\geq 1/2$ or $\theta\leq
-1/2$. So hereafter, we only discuss the behaviors for $-1/2<\theta<1/2$
so that $A$ remains finite. $S(q)$ shows the power low behavior for
$q\ll \mu^{1/2}\approx \xi^{-1}$:
\begin{align}
 S(q) \sim q^{-2\rho}.
\end{align}
For $\rho>0$, $S(q)\to \infty$ for small $q$, 
leading to the power-low correlation
\begin{align}
 G(\bx)=\ave{\phi(\bx)\phi(0)}\sim \abs{\bx}^{2\rho-d}.\label{184115_23Feb23}
\end{align}
As a consequence, the fluctuation of the order parameter in the
$d$-dimensional square box $[0,l]^d$ behaves
as~\cite{torquato2018hyperuniform}
\begin{align}
\sigma(l)^2\equiv \ave{\left(\int_{\bx\in
[0,l]^d} d\bx \phi(\bx)\right)^2}\sim
 l^{d+2\rho},
\label{093322_17Mar23}
 \end{align}
which is much larger than the naive expectation from the central limit
theorem $\sigma^2\sim l^d$. This anomalous enhancement of the
fluctuation is the signature of the GNF~\cite{narayan2007long}. For
$\rho<0$, $S(q)\to 0$ in the limit $q\to 0$. In this case, the
fluctuation of the order parameter Eq.~(\ref{093322_17Mar23}) is highly
suppressed, {\it i.e.}, the model exhibits the
HU~\cite{torquato2018hyperuniform}. To visualize these results, we show
typical behaviors of $S(q)$ in Fig.~\ref{091838_2Aug23}.

\subsection{Lagrange multiplier}
The remaining task is to determine the Lagrange multiplier $\mu$ by the
spherical constraint:
\begin{align}
N=\int d\bx \ave{\phi(\bx,t)^2}
  = \frac{V}{(2\pi)^d}\int d\bq S(q),\label{042636_17Mar23}
\end{align}
where $V=\int d\bm x$ denotes the volume of the system.
Substituting Eq.~(\ref{sq}) into
Eq.~(\ref{042636_17Mar23}), we get
\begin{align}
1 = TA'\int_0^{q_D}dq\frac{q^{d-1-2\rho}}{(q^2+\mu)^{1+2\theta}},\label{155550_19Feb23}
\end{align}
where $q_D$ denotes the cut-off and 
\begin{align}
 A' = \frac{\Omega_d A}{(2\pi)^d}\frac{V}{N}.
\end{align} 
Here $\Omega_d$ denotes the $d$ dimensional solid angle. Substituting
$\mu=0$ into Eq.~(\ref{155550_19Feb23}), one can calculate the critical
temperature $T_c$ as follows:
\begin{align}
T_c = 
\begin{cases}
 0 & d\leq d_l\\
 (d-d_l)/A'q_D^{d-d_l}& d >d_l
\end{cases},\label{111951_10Mar23}
\end{align}
where  we have defined the lower critical dimension as 
\begin{align} 
 d_l = 2+2\rho+4\theta,\ (-1/2<\theta<1/2).
\end{align}
This is consistent with the result of the dimensional analysis
Eq.~(\ref{024537_17Mar23}) for $\rho>-1/2$. On the contrary, the results
are inconsistent for $\rho<-1/2$. Further studies would be beneficial to
elucidate this point, but anyway, the qualitative result remains the
same: the positive correlation increases $d_l$, and the anticorrelation
reduces $d_l$.

The detailed analysis of Eq.~(\ref{155550_19Feb23}) near $T_c$
leads to (see Appendix.~\ref{123702_19Aug22})
\begin{align}
\mu \sim (T-T_c)^\gamma
\end{align}
with
\begin{align}
 \gamma =
 \begin{cases}
  \frac{2}{d-d_l} & d_l< d < d_u \\
  1 & d > d_u,
 \end{cases},
\end{align}
where the upper critical dimension $d_u$ is
\begin{align}
d_u = 4+2\rho+4\theta, (-1/2<\theta<1/2).\label{093620_21Feb23}
\end{align}
Again the result is consistent with the dimensional analysis for
$\rho>-1/2$.  Substituting this result into Eq.~(\ref{160435_18Feb23}),
we can determine the critical exponent:
\begin{align}
 \xi \sim (T-T_c)^{-\nu}\label{172303_20Feb23}
\end{align}
with
\begin{align}
 \nu =
 \begin{cases}
  1/(d-d_l) & d_l< d < d_u \\
  1/2 & d > d_u
 \end{cases}.
\end{align}
The critical exponent differs from the equilibrium value if
$2\rho+4\theta\neq 0$ since $d_l\neq d_l^{\rm eq}=2$.  In other words,
the long-range spatio-temporal correlation of the noise changes the
universality class.

\section{Conserved order parameter}
\label{145218_19Mar23}

Let $\vec{\phi}=\{\phi_1,\cdots,\phi_n\}$ be a conserved $n$-component order parameter.  We
assume that the time evolution of $\phi_a(\bx,t)$ is described by the
model-B dynamics~\cite{chaikin1995principles}:
\begin{align}
\pdiff{\phi_a(\bx,t)}{t} =
 \Gamma\nabla^2\fdiff{F[\vec{\phi}]}{\phi_a(\bx,t)}
 + \nabla\cdot\bm{\xi}_a(\bx,t),\label{094852_17Mar23}
\end{align}
where $\Gamma$ denotes the damping coefficient,
${\bm{\xi}_a=\{\xi_{a,\mu}\}_{\mu=1,\dots,d}}$ denotes the noise, and $d$ denotes
the spatial dimension. The mean and variance of the noise $\xi_{a,\mu}(\bx,t)$ are given
by
\begin{align}
&\ave{\xi_{a,\mu}(\bx,t)} =0,\new 
&\ave{\xi_{a,\mu}(\bx,t)\xi_{b,\nu}(\bx',t')} =2T\delta_{ab}\delta_{\mu\nu}\Gamma D(\bx-\bx',t-t'),\label{135947_27Jul23}
\end{align}
where the Fourier transform of $D(\bx,t)$ is given by
Eq.~(\ref{094611_17Mar23}).

\subsection{Dimensional analysis for $O(n)$ model}
\label{dim}

Substituting the free energy Eq.~(\ref{153105_27Feb23}) into
Eq.~(\ref{094852_17Mar23}), we get
\begin{align}
\dot{\phi}_a= \Gamma\nabla^2 (-\nabla^2\phi_a+\varepsilon\phi_a + g\phi_a|\vec{\phi}|^2)+\nabla\cdot\bm{\xi}_a.\label{155729_20Mar23}
\end{align}
As before, we consider the scaling transformations: $x\to bx$, $t\to
b^{z_t}t$, $\phi_a\to b^{z_\phi}\phi_a$, $g\to
b^{z_g}g$~\cite{nishimori2010elements}. Assuming the scaling invariance
of the dynamic equation Eq.~(\ref{155729_20Mar23}), we get
\begin{align}
&z_t = 4,\new
&z_g = 2-2z_\phi-z_t,\new
&z_\phi =
1 + \frac{2\rho'-d}{2}+4\theta',
\end{align}
where $\rho'=\max[\rho,-1/2]$ and $\theta'=\max[\theta,-1/2]$, as
defined in Eq.~(\ref{170358_1Aug23}).  As before, the lower critical
dimension is calculated by setting $z_{\phi}=0$, leading to
\begin{align}
d_l =
2+2\rho'+8\theta'.
\end{align}
The upper critical dimension is obtained by setting $z_g=0$, leading to
\begin{align}
d_u =
4+2\rho'+8\theta'.
\end{align}
When $\theta=0$, the results are consistent with those of the model-A,
see Sec.~\ref{173631_21Feb23}, while when $\theta\neq 0$, we get
different results. Aside from such a difference, the qualitative
conclusion remains the same: the positive correlation of the noise
($\rho>0$ and $\theta>0$) increases the critical dimensions $d_l$ and
$d_u$, while the anticorrelation ($\rho<0$ and $\theta<0$) reduces
$d_l$ and $d_u$.

\subsection{Spherical model}

\begin{figure}[t]
\begin{center}
\includegraphics[width=9cm]{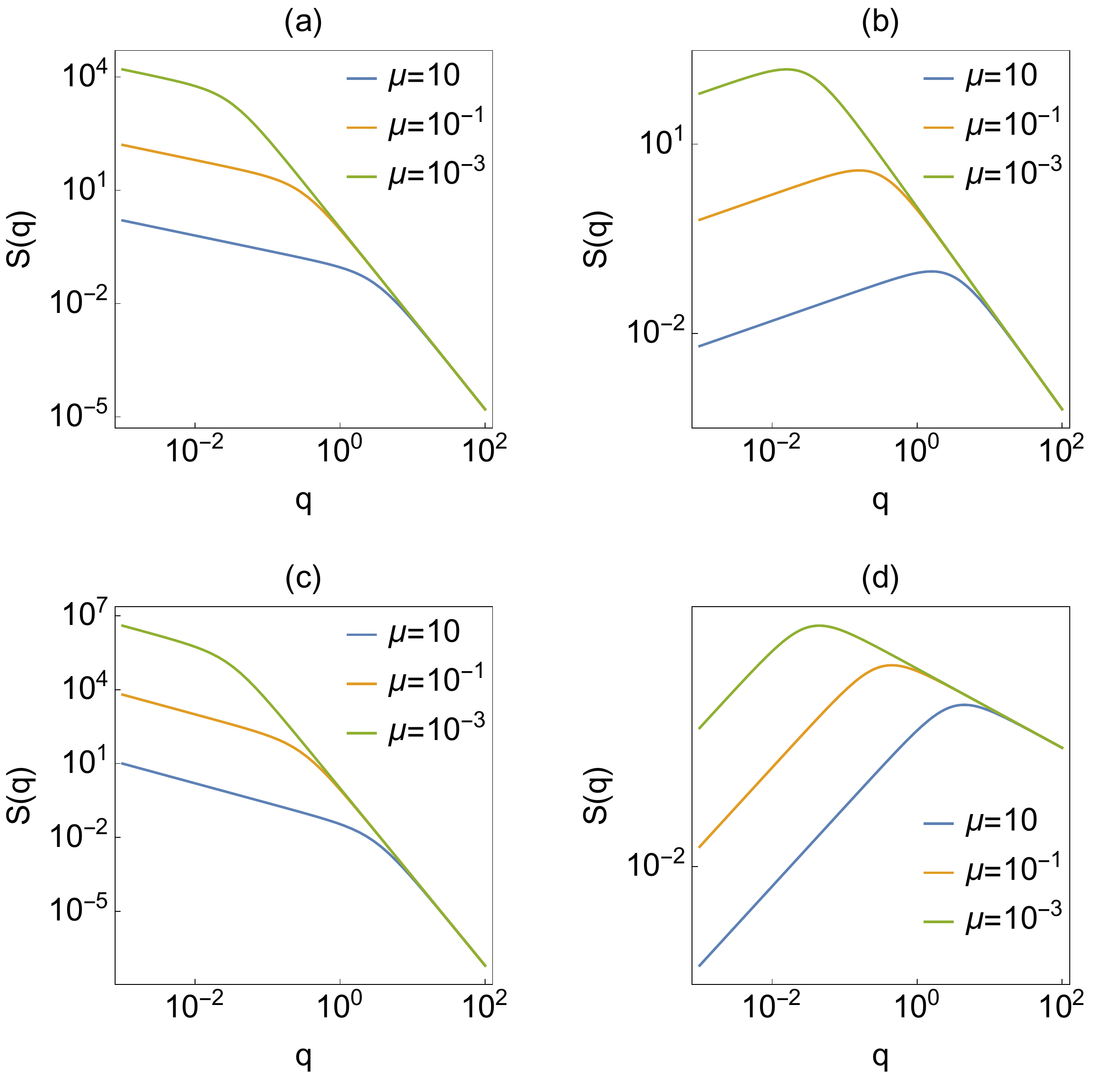} \caption{Typical behaviors of $S(q)$
of spherical model for model-B: (a) $S(q)$ for positive spatial
correlation ($\rho=0.2, \theta=0$).  (b) $S(q)$ for negative spatial
correlation ($\rho=-0.2, \theta=0$). (c) $S(q)$ for positive temporal
correlation ($\rho=0, \theta=0.2$).  (d) $S(q)$ for negative temporal
correlation ($\rho=0, \theta=-0.2$).  For positive spatial or temporal
correlation, $S(q)$ diverges in the limit of small $q$ even far from the
transition point, {\it i.e.}, the GNF appears. For negative spatial or
temporal correlation, on the contrary, $S(q)$ vanishes in the limit of
small $q$, {\it i.e.}, the HU appears. For simplicity, here we set
$B=T=1$.}  \label{111144_2Aug23}
\end{center}
\end{figure}

The spherical model for the model-B dynamics is 
\begin{align}
\dot{\phi}= \Gamma\nabla^2 (-\nabla^2\phi+\mu\phi)+\nabla\cdot\bm{\xi}.
\end{align}
This model neglects the non-linear term of the $O(n)$ model and instead
impose the spherical constraint ${\int d\bx\ave{\phi^2} = N}$.  One
can solve it easily since this is a linear equation.  For instance, the
static structure factor $S(\bq)$ in the steady-state is calculated as
\begin{align}
S(\bq)& = \frac{BTq^{-2\rho-4\theta}}{(q^2+\mu)^{1+2\theta}}\label{215402_19Feb23}
\end{align}
where $B$ denotes a constant.
\begin{align}
B = \frac{1}{\Gamma^{2\theta}\pi}\int_{-\infty}^\infty \frac{\abs{x}^{-2\theta}dx}{x^2+1}
= \frac{\sec(\pi\theta)}{\Gamma^{2\theta}}.
\end{align}
As before, we restrict $\theta$ to $-1/2<\theta<1/2$ to keep $B$ finite.
$S(q)$ shows the power low behavior for $q\ll \mu^{1/2}\approx \xi^{-1}$:
\begin{align}
 S(q) \sim q^{-2\rho-4\theta}.\label{134827_27Jul23}
\end{align}
For $2\rho+4\theta>0$, $S(q)\to \infty$ for small $q$, leading to the
GNF~\cite{narayan2007long}. On the contrary, for $2\rho+4\theta<0$,
$S(q)\to 0$ for small $q$, leading to the
HU~\cite{torquato2018hyperuniform}. Interestingly, the GNF and HU appear
even without the spatial correlation of the noise $\rho=0$.  To
visualize the above results, we show typical behaviors of $S(q)$ in
Fig.~\ref{111144_2Aug23}.

The Lagrange multiplier $\mu$ is to be determined by the spherical
constraint ${\int d\bx \ave{\phi^2}=N}$. As before, the detailed
analysis of this equation allows us to calculate the lower and upper
critical dimensions for ${-1/2<\theta<1/2}$ (see
Appendix.~\ref{123702_19Aug22}):
\begin{align} 
&d_l = 2+2\rho+8\theta,\new
&d_u = 4+2\rho+8\theta.
\end{align}
For $\rho>-1/2$, the results are consistent with the dimensional
analysis in the previous subsection. On the contrary, for $\rho<-1/2$,
we get inconsistent results. Further studies would be beneficial to
elucidate the origin of this discrepancy. Aside from such a minor
difference, both $O(n)$ and spherical models predict that the
positive correlation of the noise ($\rho>0$ and $\theta>0$) increases
the critical dimensions, $d_l$ and $d_u$, while the anticorrelation
reduces $d_l$ and $d_u$.

For $d_l<d<d_u$ near $T_c$, the scaling behaviors of the
Lagrange multiplier $\mu$, 
correlation length $\xi$, and relaxation time $\tau$
are (see Appendix.~\ref{123702_19Aug22})
\begin{align}
&\mu\sim (T-T_c)^\gamma,&
&\xi \sim (T-T_c)^\nu,&
&\tau \sim \xi^z,
\end{align}
where
\begin{align}
&\gamma = \frac{2}{d-d_l},&
&\nu = \frac{1}{d-d_l},&
&z =4.
\end{align}
Note that the static critical exponent $\nu$ differs from the 
equilibrium values
if $2\rho+8\theta\neq 0$ because $d_l\neq d_l^{\rm eq} \equiv
2$~\cite{berlin1952spherical,nishimori2010elements}. This implies that
the long-range spatio-temporal correlation of the noise leads to a new
universality class. Further theoretical and numerical studies would be
beneficial to elucidate this point.

\subsection{Center of mass conserving dynamics}
An interesting application is for the systems driven by the center of
mass conserving (COMC) dynamics, which was introduced to explain the
HU~\cite{hexner2017noise,lei2019hydrodynamics}. The origin
of the HU may depend on the detail of the system, but
Hexner and Levine claimed that there is a universal mechanism to yield
HU for systems conserving the center of
mass~\cite{hexner2017noise}. Below we briefly summarize their argument
for the model-B dynamics. Let $\rho(\bx,t)$ be the density following the
equation of continuity:
\begin{align}
\pdiff{\rho(\bx,t)}{t} = -\nabla\cdot\bm{J(\bx,t)},\label{133243_27Jul23}
\end{align}
where $\bm{J}$ denotes the flux.
In the case of the standard model-B,
$\bm{J}$ is written as~\cite{onuki2002phase}
\begin{align}
\bm{J}= -\Gamma\nabla\mu
 + \bm{\xi},\label{133253_27Jul23}
\end{align}
where $\mu=\delta F/\delta \rho(\bx,t)$ denotes the chemical potential and $\bm{\xi}$ denotes the
noise. The conservation of the center of mass
requires~\cite{hexner2017noise}
\begin{align}
\diff{\ave{x_a}}{t}&=\int d\bx x_a\pdiff{\rho(\bx,t)}{t}\new
 &= -\int d\bx x_a\nabla\cdot\bm{J}(\bx,t)\new
 &= \int d\bx J_a(\bx,t) \new
 &= \int d\bx \xi_a = 0.
\end{align}
To satisfy the last equality, $\xi_a$ should be written in the form of 
a divergence of other vector:
\begin{align}
\xi_a(\bx,t) = \nabla\cdot\bm{\sigma}_a(\bx,t) = \sum_{b}\pdiff{\sigma_{ab}(\bx,t)}{x_b}.\label{133301_27Jul23}
\end{align}
The simplest choice of $\sigma_{ab}(\bx,t)$ is an isotropic white noise:
\begin{align}
\ave{\sigma_{ab}(\bx,t)\sigma_{cd}(\bx',t')}\propto \delta_{ac}\delta_{bd}\delta(\bx-\bx')\delta(t-t').
\end{align}
Then we get 
\begin{align}
 \ave{\xi_a(\bx,t)\xi_b(\bx',t')}\propto \delta_{ab}\nabla^2\delta(\bx-\bx')\delta(t-t'),
\end{align}
which is tantamount to set $D(\bq,\omega)\propto q^2$ in our model-B
dynamics, {\it i.e.}, $\rho=-1$ and $\theta=0$, see
Eq.~(\ref{135947_27Jul23}). In this case, the static structure factor
Eq.~(\ref{134827_27Jul23}) behaves as $S(q)\sim q^2$ for a small waver
number $q$. Therefore, the model exhibits the HU, as discussed in
previous work~\cite{hexner2017noise}.  Also, the dimensional
analysis of the $O(n)$ model and spherical model both predict that the
lower critical dimension becomes lower than the equilibrium value,
$d_l<d_l^{\rm eq}=2$. This implies that the continuous symmetric
breaking can occur even in $d=2$ or lower dimension in contrast with the
equilibrium systems for which the Mermin-Wagner theorem prohibits the
long-range order. This appears to be consistent with a recent numerical
simulation of a two-dimensional system driven by the COMC dynamics,
where the authors reported the emergence of the perfect crystal phase
even in $d=2$~\cite{leonardo2023}. But strictly speaking, our model is
for the second-order phase transition, and thus it can not be directly
applied to the first-order phase transition such as the
crystaillization. Further numerical and theoretical studies would be
beneficial.

\section{Summary and discussions}
\label{130256_22Feb23}

\begin{table}[h]
\centering
\label{tab:modelA}
\begin{tabular}{|c|c|c|}
\hline
Model-A & $d_l$ & $d_u$ \\
\hline
$O(n)$ model & $2+2\rho'+4\theta'$ & $4+2\rho'+4\theta'$ \\
Spherical model $(-1/2<\theta<1/2)$ & $2+2\rho+4\theta$ & $4+2\rho+4\theta$ \\
\hline
Model-B & $d_l$ & $d_u$ \\
\hline
$O(n)$ model & $2+2\rho'+8\theta'$ & $4+2\rho'+8\theta'$ \\
Spherical model $(-1/2<\theta<1/2)$& $2+2\rho+8\theta$ & $4+2\rho+8\theta$ \\
\hline 
\end{tabular}
\caption{Critical dimensions. Here we used abbreviations
$\rho'=\max[-1/2,\rho]$ and $\theta'=\max[-1/2,\theta]$.}
\label{170647_1Aug23}
\end{table}

\begin{figure}[t]
\begin{center}
\includegraphics[width=8.5cm]{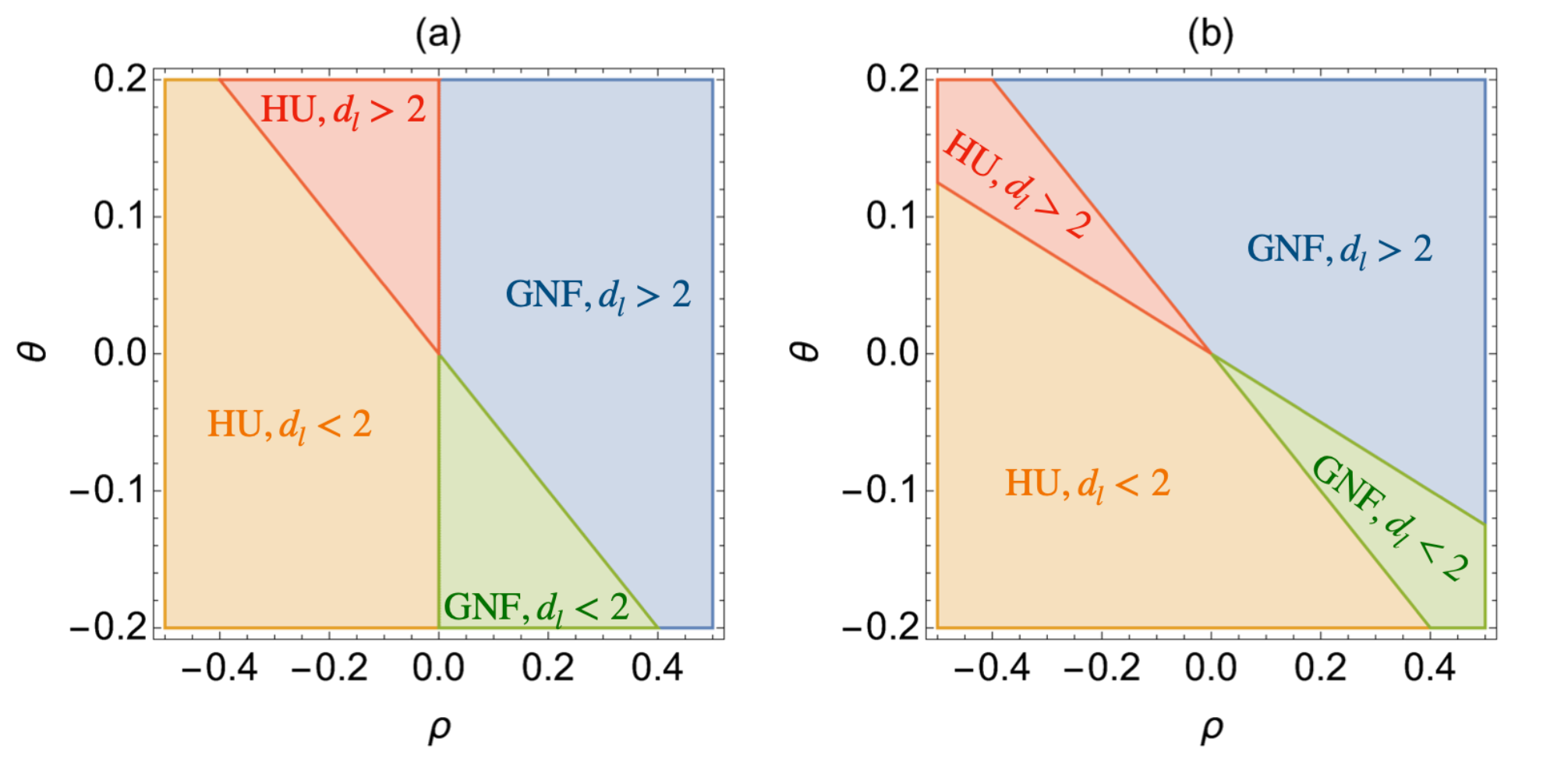} \caption{Phase behaviors for
the model-A (a) and model-B (b).}  \label{123150_26Jul23}
\end{center}
\end{figure}

\subsection{Summary}
In this work, we calculated the lower and upper critical dimensions,
$d_l$ and $d_u$, of the $O(n)$ and spherical models driven by the
model A and B dynamics with the correlated noise $\xi(\bx,t)$, see
Table.~\ref{170647_1Aug23} for a summary. The correlation of the noise
is written in the Fourier space as $D(\bq,\omega) =
\abs{\bq}^{-2\rho}\abs{\omega}^{-2\theta}$. Our results imply that the
positive correlation of the noise ($\rho>0$ and $\theta>0$) increases
the critical dimensions, $d_l$ and $d_u$, while the anticorrelation 
($\rho<0$ and $\theta<0$) reduces $d_l$ and $d_u$. We also found that
the static structure factor $S(q)$ in the paramagnetic phase exhibits
the power-low behavior for small wave number $S(q)\sim q^\alpha$ with
$\alpha=-2\rho$ for the non-conserved order parameter (model-A) and
$\alpha=-2\rho-4\theta$ for the conserved order parameter (model-B),
leading to the GNF for $\alpha<0$, and HU for $\alpha>0$. We summarize
those results in Fig.~\ref{123150_26Jul23}.

\subsection{Hyperuniformity, gian number fluctuation, and lower critical dimension}
Our first expectation was that the HU would decrease $d_l$ from the
equilibrium value $d_l^{\rm eq}=2$, and the GNF would increase
$d_l$. This expectation is correct for the spatiotemporally positive
correlated noise ($\rho>0, \theta>0$) and anticorrelated noise ($\rho<0,
\theta<0$), see Fig.~\ref{123150_26Jul23}. However, for the intermediate
cases, $\rho<0, \theta>0$ and $\rho>0, \theta<0$, the HU or GNF does not
always guaranty $d_l<2$ or $d_l>2$, see
Fig.~\ref{123150_26Jul23}. Further studies would be beneficial to
elucidate the relation between the HU, GNF, and $d_l$.

\subsection{Perspective on temporally correlated noise}
The temporally correlated noise for $-1/2<\theta<1/2$ has been studied
extensively in the context of the anomalous diffusion in crowded
environments, because a free particle driven by the noise $\dot{x}=\xi$
exhibits the sub-diffusion $\ave{x(t)^2}\sim t^{1+2\theta}$ for
$-1/2<\theta<0$ and super-diffusion for $0<\theta<1/2$, see
Refs.~\cite{bouchaud1990anomalous,hofling2013anomalous} for reviews.
However, relatively few studies have been done on the effects of the
temporal correlation on critical phenomena. For example, in
Refs.~\cite{garc1994,sancho1998non,maggi2022critical}, the authors
studied the effect of exponentially correlated noise on the $\varphi^4$
model and found the same universality as the equilibrium Ising model. In
Ref.~\cite{bonart2012critical}, the authors studied the $O(n)$ model
with the power-low correlated noise, but the noise was introduced in a
way that preserves the detailed balance. Thus, the critical dimensions
and the static critical exponents are unchanged from those in
equilibrium.  On the contrary, non-equilibrium noises, such as the $1/f$
noises, often show the power-low frequency dependence of the power
spectrum, naturally leading to the long-range temporal
correlation~\cite{richard1976,per1987,dutta1981,milotti2002,eliazar2009unified}.
Our research has demonstrated the emergence of novel phenomena, such as
the GNF, HU, and new universality classes
in systems driven by such long-range temporally correlated noise. We
hope that our findings will motivate further investigation into the
fascinating properties of these systems.

\subsection{Systems driven by imperfect periodic or quasiperiodic forces}
\label{imper}

Tissues are often driven by periodic deformation of
cells~\cite{zhang2022pulsating}.  In chiral active matter, constituent
particles spontaneously rotate due to asymmetry of the driving
forces~\cite{liebchen2022chiral,kuroda2023microscopic}. The driving
force of those systems would be approximated by temporally periodic
functions. In previous work, we investigated a model driven by
temporally periodic but spatially uncorrelated driving forces and found
that the model exhibits the HU and smaller value of the lower critical
dimension than that in equilibrium $d_l<2$~\cite{ikeda2023does}.
However, the completely periodic function does not exist in reality due
to friction or other uncontrollable effects. The effects of the imperfection
of periodic patterns have been investigated extensively in the context
of the HU, and it is known that the Fourier spectrum often exhibits the
power-low with a positive exponent in these
cases~\cite{kim2018,uz2019hyperuniformity}. For the simplest example, we
consider an imperfect periodic pulse~\cite{kim2018}:
\begin{align}
\xi(t) = \lim_{T\to\infty}\sum_{n=1}^T\delta(t-na-\eta_n),\label{113754_30Jul23}
\end{align}
where $\eta_n$ represents a perturbation to a periodic 
pulse. For simplicity, let us assume that $\eta_n$ is an i.i.d Gaussian
random variable of zero mean and variance $\sigma$.  Then, the power
spectrum of $\xi(t)$ can be calculated as follows~\cite{kim2018}
\begin{align}
D(\omega) &= \lim_{T\to\infty}\overline{\frac{1}{T}\abs{\sum_{n=1}^T e^{i\omega(na+\eta_n)}}^2}\new 
 &= 1 + e^{-\sigma\omega^2} (D_0(\omega)-1),
\end{align}
where the overline denotes the average for $\eta_n$, and $D_0(\omega)$
represents the spectrum of the periodic pulse:
\begin{align}
D_0(\omega) = \lim_{T\to\infty}\frac{1}{T}\abs{\sum_{n=1}^T e^{i\omega na}}^2.
\end{align}
$D_0(\omega)$ is nothing but the static structure factor of a
one-dimensional lattice, and in particular $D_0(\omega)=0$ for
sufficiently small $\omega$~\cite{kim2018}. For $\omega\ll 1$, we get
\begin{align}
 D(\omega)\sim 1-e^{-\sigma\omega^2}\sim \sigma\omega^{-2\theta}
\end{align}
with $\theta=-1$. This simple example demonstrated that the power-law
spectrum with negative $\theta$ can naturally arise due to the
imperfection of the periodic pattern. More systematic studies for
various types of imperfections have been investigated in
Ref.~\cite{kim2018}. The similar power law of the Fourier spectrum has
been also reported for one-dimensional quasi-periodic
sequences~\cite{oguz2017,uz2019hyperuniformity}.  Do systems driven by
imperfect periodic or quasi-periodic forces exhibit the HU and symmetry
breaking transition in $d\leq 2$, as predicted by our theory?  Further
theoretical and numerical studies would be beneficial to elucidate this
point.

\acknowledgments We thank K.~Miyazaki and A.~Ikeda for useful
discussions. This project has received JSPS KAKENHI Grant Numbers
21K20355 and 23K13031.

\appendix

\section{Scaling of $\mu$}
\label{123702_19Aug22}

To determine $\mu$, one should solve the following self-consistent equation:
\begin{align}
1 = TG(\mu)\equiv TA\int_0^{q_D} dq  \frac{q^{d-1+m}}{(q^2+\mu)^n},\label{122931_12Aug22}
\end{align}
where $A$, $n$, and $m$ are constants.
We want to derive the scaling behavior of $\mu$ near the critical point:
\begin{align}
 T_c = \left[A\int_0^{q_D} dqq^{d-1+m-2n}\right]^{-1}.\label{153040_18Feb23}
\end{align}
For $d+m-2n>0$, the denominator of Eq.~(\ref{153040_18Feb23}) diverges,
and thus the model does not have the critical point at finite $T$.
This implies that the lower critical dimension is 
\begin{align}
d_l = 2n-m.
\end{align}
When $d>2n-m+2$, $G(\mu)$ can be expanded as 
\begin{align}
\frac{1}{T} &= G(0) + \mu G'(0) + \cdots\new 
 &= \frac{1}{T_c} + \mu G'(0) + \cdots,
\end{align}
leading to
\begin{align}
\mu \sim (T-T_c)^{1}.
\end{align}
On the contrary, if $d\in (2n-m,2n-m+2)$, $G'(\mu)$ for small $\mu$ 
behaves as 
\begin{align}
 G'(\mu) \sim \mu^{\frac{d+m-(2n+2)}{2}},\label{120735_13Aug22}
\end{align}
implying 
\begin{align}
 G(\mu) -G(0) =\int_0^{\mu}d\mu' G'(\mu') \sim \mu^{\frac{d+m-2n}{2}},
\end{align}
leading to
\begin{align}
 \frac{1}{T} = G(\mu) = \frac{1}{T_c}-B\mu^{\frac{d+m-2n}{2}},
\end{align}
where $B$ is a constant. Therefore, the scaling of $\mu$ for $\mu\ll 1$ is 
\begin{align}
\mu \sim (T-T_c)^{\frac{2}{d+m-2n}} \sim (T-T_c)^{\frac{2}{d-d_l}}.
\end{align}
The above results imply that the
upper critical dimension is
\begin{align}
 d_u = 2n-m+2 = d_l+2.
\end{align}

\bibliography{reference}

\end{document}